\documentclass[12pt,epsf,amssymb]{article}

\usepackage{graphicx}
\usepackage{axodraw}

\begin{document}
\baselineskip 0.6cm
\newcommand{\gsim}{ \mathop{}_{\textstyle \sim}^{\textstyle >} }
\newcommand{\lsim}{ \mathop{}_{\textstyle \sim}^{\textstyle <} }
\newcommand{\vev}[1]{ \left\langle {#1} \right\rangle }
\newcommand{\bra}[1]{ \langle {#1} | }
\newcommand{\ket}[1]{ | {#1} \rangle }
\newcommand{\Dsl}{\mbox{\ooalign{\hfil/\hfil\crcr$D$}}}
\newcommand{\nequiv}{\mbox{\ooalign{\hfil/\hfil\crcr$\equiv$}}}
\newcommand{\EV}{ {\rm eV} }
\newcommand{\KEV}{ {\rm keV} }
\newcommand{\MEV}{ {\rm MeV} }
\newcommand{\GEV}{ {\rm GeV} }
\newcommand{\TEV}{ {\rm TeV} }

\def\diag{\mathop{\rm diag}\nolimits}
\def\tr{\mathop{\rm tr}}

\def\Spin{\mathop{\rm Spin}}
\def\SO{\mathop{\rm SO}}
\def\O{\mathop{\rm O}}
\def\SU{\mathop{\rm SU}}
\def\U{\mathop{\rm U}}
\def\Sp{\mathop{\rm Sp}}
\def\SL{\mathop{\rm SL}}


\begin{titlepage}

\begin{flushright}
UT-04-04 \\
LBNL-54659 \\
\end{flushright}

\vskip 2cm
\begin{center}
{\large \bf Super- and CP-symmetric QCD in Higher Dimensions}

\vskip 1.2cm
Izawa~K.-I.$^1$, Taizan Watari$^2$, and T.~Yanagida$^1$

\vskip 0.4cm
{$^1$\it Department of Physics, University of Tokyo, \\
          Tokyo 113-0033, Japan}\\
{$^2$\it Department of Physics, University of California, \\
          Berkeley, CA 94720, USA}\\
\vskip 1.5cm
\abstract{
An extremely precise global symmetry is necessary 
in the Peccei--Quinn solution to the strong CP problem. 
Such symmetry arises when colored chiral fermions are 
localized in an internal space. We present a supersymmetric model 
that incorporates the above mechanism. 
Extra colored chiral multiplets around the supersymmetry-breaking 
scale are a generic prediction of the supersymetric model. }

\end{center}
\end{titlepage}

\setcounter{page}{2}


\section{Introduction}

The Peccei--Quinn mechanism \cite{Pec} is a promising solution 
to the strong CP problem, yet it requires an extremely precise 
global symmetry. The explicit breaking of the symmetry 
gives rise to an extra scalar potential of the axion. 
The energy scale relevant to the mechanism 
has to be $10^{10}$--$10^{12}$ GeV, while 
the extra potential has to be $10^{-10}$ times smaller than 
the potential generated by the QCD dynamics. 
What is the origin of such a precise global symmetry?

Suppose that the SU(3)$_C$ gauge field propagates in extra 
space dimensions, and that some of SU(3)$_C$-charged 
chiral fermions are localized in the internal space.
Then, an approximate chiral symmetry naturally arises when 
the chiral fermions are separated sufficiently 
in the internal space.
Such a symmetry might well be stable even against 
possible quantum gravitational corrections.
Our previous papers \cite{Iza} identified it with the Peccei--Quinn symmetry.

Supersymmetry is a leading candidate to be
discovered through experiments in no distant future.
As such, it is tempting to extend the higher-dimensional
QCD, which naturally realizes the strong CP invariance
in the nonsupersymmetric case
mentioned above,
to a supersymmetric setup.
In this article, we present an explicit model, and show that 
colored chiral multiplets are around the supersymmetry-breaking 
scale in addition to the particles of the minimal supersymmetric
standard model (MSSM).

\section{Accidental Axial Symmetry}

In this section, we provide a basic supersymmetric setup 
for the accidental axial (Peccei--Quinn) symmetry.

\subsection{Bulk color gauge theory}

Let us consider four-dimensional Minkowski spacetime $M_4$ 
along with one-dimensional extra space $S^1$, whose coordinate $y$ 
extends from $-l$ to $l$ (that is, two points at $y=l$ and $y=-l$ 
are identified). The SU(3)$_C$ vector field is assumed to 
propagate on the whole spacetime $M_4\times S^1$. Thus, 
the SU(3)$_C$ vector field belongs to a vector multiplet $\Psi$ 
of the minimal supersymmetry (SUSY) in five dimensions. 
One vector multiplet $\Psi$ consists of 
one vector multiplet $V$ and one 
chiral multiplet $\Phi $ of ${\cal N}$ = 1 SUSY 
in four-dimensional spacetime.

The minimal SUSY in five-dimensional spacetime corresponds to 
${\cal N}$ = 2 SUSY in four dimensions. 
Thus, description of gauge theories based on 
four-dimensional ${\cal N}$ = 2 SUSY
is useful
in considering SUSY theories in five dimensions.
The action of a U(1) vector multiplet $\Psi=(V, \Phi)$
of ${\cal N}$ = 2 in four dimensions is given in terms of 
a holomorphic function ${\cal F}(\Psi)$:
\begin{eqnarray}
K = \frac{1}{4\pi}{\rm Im}\left( \frac{\partial {\cal F}(\Phi)}
{\partial \Phi}\Phi^ \dagger \right), \quad
W = \frac{-i}{16\pi} 
    \left( \frac{\partial^2 {\cal F}(\Phi)}{\partial \Phi^2} 
        \right)
        {\cal W}^\alpha {\cal W}_\alpha,
\end{eqnarray}
where $K$ and $W$ denote the K{\" a}hler potential and superpotential,
respectively.

In five-dimensional spacetime, the imaginary part
of the scalar component in $\Phi$ 
is the 4-th polarization of the five-dimensional vector field.
The superfields $(V(x),\Phi(x))$ are promoted to fields 
in five dimensions $(V(x,y),\Phi(x,y))$. 
The K{\" a}hler potential and superpotential are now integrated over 
$d^4x dy d^2\theta d^2\bar{\theta}$ and $d^4x dy d^2\theta$, 
respectively.
The coordinate $y$ is integrated along $S^1$. 
Let us consider a prepotential that is an at-most-cubic polynomial 
\begin{equation}
{\cal F} = i\frac{1}{2}\frac{4 \pi}{g^2}\Psi^2 + i h\Psi^3.
\end{equation}
The quadratic term provides the gauge kinetic term, and the cubic 
term contains the Chern--Simons term
\begin{equation}
 - \int \frac{3}{\sqrt{2}8 \pi} h A_4 F_{\mu \nu} F_{\kappa \lambda}
 \epsilon^{\mu\nu\kappa\lambda} d^4x dy ,
\label{eq:CS-U(1)}
\end{equation}
where $A_4$ denotes the 4-th polarization of the gauge field and 
$F_{\mu\nu}$ the field strength in four dimensions.
The Lorentz symmetry and gauge covariance can be
restored by modifying the action
\cite{Sei,Ark}.
The precise form of the Chern--Simons term in five dimensions 
in the case of nonabelian gauge group is found in Ref.\cite{Cer} 
(with supergravity).
We now turn to the SU(3)$_C$ gauge group.

Kaluza--Klein reduction to the four-dimensional spacetime, however, 
yields an unwanted ${\cal N}$ =1 chiral multiplet
in the SU(3)$_C$-{\bf adj.} representation
in the low energy spectrum, and 
the ${\cal N}$ = 2 SUSY is left unbroken.
Hence, we consider an $S^1/Z_2$ orbifold instead of $S^1$.
The multiplets $V(x,y)$ and $\Phi(x,y)$ of ${\cal N}$ = 1 SUSY are 
now under a constraint 
\begin{equation}
V(x,y) = V(x,-y), \quad \Phi(x,y) = - \Phi(x,-y).
\end{equation}
Then, we have only an SU(3)$_C$ vector multiplet of the 
${\cal N}$ = 1 SUSY without the chiral multiplet $\Phi$ 
after the Kaluza--Klein reduction.
In order to define the theory consistently, the action in 
$M_4 \times S^1$ should be invariant under the $Z_2$ transformation.
The invariance is achieved as long as 
i) the coefficient of the quadratic term $1/g^2$ 
is even under the $Z_2$ transformation, and 
ii) that of the cubic term $h$ is odd.
We take $1/g^2$ to be $y$-independent and $h(y)$ as 
\begin{equation}
h(y) = c \frac{y}{|y|},
\label{eq:kink}
\end{equation}
where $c$ is a constant
that plays the same role as the corresponding one
in the non-SUSY models
\cite{Iza}(see below).%
\footnote{
In the appendix,
the parameter $h(y)$ is interpreted as a vacuum expectation value 
of a background field, 
where a possible origin of the kink configuration is also discussed.}

\subsection{Boundary extra quarks}

There are two fixed points in the $S^1/Z_2$ orbifold: 
$y=0$ and $y=l$. Let us put SU(3)$_C$-charged chiral multiplets 
as extra quarks on the fixed-point boundaries: 
chiral multiplets $Q$ in the SU(3)$_C$-{\bf 3} 
representation at $y=0$ 
and the same number of chiral multiplets $\bar{Q}$
in the SU(3)$_C$-{\bf 3}$^*$ representation at $y=l$. 
The action of the extra-quark multiplets contains 
\begin{equation} 
\int_{y=0} d^4x d^2\theta d^2\bar{\theta} Q^* e^V Q 
+ \int_{y=l} d^4 x d^2\theta d^2\bar{\theta}
   \bar{Q} e^{-V}\bar{Q}^*.
\end{equation}

These chiral multiplets by themselves
give rise to a triangle anomaly of SU(3)$_C$ 
at each fixed point. However, the anomaly at $y=0$ is the same as that 
at $y=l$ with the opposite sign. Thus, the anomaly can be
canceled through 
its flow between $y = 0$ and $y=l$ implemented by the 
Chern--Simons interaction (\ref{eq:CS-U(1)}), provided 
the constant $c$ is adequately chosen.\footnote{
See Ref.\cite{Iza} for numerical details.}

When the extra dimension is sufficiently large, interactions 
involving both $Q$ and $\bar{Q}$ are highly suppressed. 
Let the cutoff scale (such as the grand unification
or Planck scale) of the model be given by $M$. 
Then the effects of particles with masses of order $M$
may generically induce such terms as
\begin{equation}
 e^{-M l} MQ\bar{Q}
\label{eq:mass}
\end{equation}
in the effective superpotential.
We assume that $M l \gsim 10^2$ to suppress the effects of such terms.
Then there is an accidental axial symmetry 
\begin{equation}
 Q \rightarrow e^{i\xi} Q, \quad 
 \bar{Q} \rightarrow e^{i \xi} \bar{Q},
\end{equation}
which is to be identified with the Peccei--Quinn symmetry.
Although interactions
comprised of only $Q$'s or $\bar{Q}$'s 
are not expected to be suppressed by $e^{-M l}$, 
they are higher-dimensional operators, and are irrelevant 
to the axion potential, as we explicitly see in the next section.

The internal dimension is moderately large:
we take $M l \simeq 10^2 \mbox{--}10^4$.
It follows that $M \simeq (10^{-1}\mbox{--}10^{-2})M_{\rm pl}$, 
and $l^{-1} \simeq (10^{-3}\mbox{--}10^{-6})M_{\rm pl}$, 
where $M_{\rm pl} \simeq 2.4 \times 10^{18}$ denotes the Planck scale.
The setup described in this section reduces to 
a four-dimensional ${\cal N}$ = 1 SU(3)$_C$ gauge theory with an accidental 
axial symmetry below the Kaluza--Klein scale.

\section{\label{sec:model}The Model}

The superpotential virtually contains no terms involving both 
$Q$ and $\bar{Q}$, 
and hence the extra quarks are not forced 
to develop a chiral condensation.
Thus, another vector multiplet is introduced 
in the five-dimensional bulk, 
so that
the chiral condensation is formed dynamically, 
and that a composite axion is obtained 
through the spontaneous chiral symmetry breaking.

For definiteness,
let us introduce an SU(5)$_H$ gauge theory%
\footnote{
This corresponds to SU(3)$_H$ in
Ref.\cite{Iza}.}
in addition to the usual color SU(3)$_C$.
The extra quarks\footnote{The usual quark and lepton chiral 
superfields are also assumed to be localized at $y=0$.} 
at $y=0$ consists of chiral superfields
$Q=(Q^i{}_{\alpha}, Q^4{}_{\alpha})$
in the $({\bf 3}+{\bf 1}) \times {\bf 5}^*$ representation
under SU(3)$_C$ $\times$ SU(5)$_H$,
and their conjugates
${\bar Q}=({\bar Q}_i{}^\alpha, \bar{Q}_4{}^\alpha)$ are at $y=l$,%
\footnote{
The color singlets $Q^4{}_\alpha$ and ${\bar Q}_4{}^\alpha$
are introduced so that
the QCD axion is obtained just in 
the same way as in Ref.~\cite{Iza}.}
where $i=1,2,3$ and $\alpha=1,\cdots,5$.

The SU(5)$_H$ gauge theory has four flavors of 
extra quarks.
Thus, an effective superpotential is generated 
due to the SU(5)$_H$ interaction 
\cite{Tay}:
\begin{eqnarray}
 W_{eff}={\Lambda_H^{11} \over \det Q{\bar Q}}, 
 \label{eq:nonpert}
\end{eqnarray}
where $\Lambda_H$ denotes the dynamical scale of 
the SU(5)$_H$ gauge theory.
The run-away potential from Eq.(\ref{eq:nonpert}) is stabilized 
by supersymmetry-breaking effects such as $V=m^2|Q|^2$ 
in the scalar potential.
We assume gravity-mediated supersymmetry breaking in this article 
for simplicity.
The Peccei--Quinn scale $F_{PQ}$ is of the order of 
\begin{eqnarray}
 \sqrt{\vev{Q{\bar Q}}} 
 \sim \left({\Lambda_H^{11} \over m}\right)^{1 \over 10}.
\end{eqnarray}

The spectrum below the Peccei--Quinn scale $F_{PQ}$ 
consists of chiral multiplets 
in the SU(3)$_C$-({\bf adj.}+{\bf 3}+{\bf 3}$^*$) representations 
and two singlets, 
in addition to the particle contents of the MSSM.
All the fermions and real scalars in the extra multiplets 
acquire masses of the order of the supersymmetry-breaking scale $m$.
Pseudo-scalars in the SU(3)$_{C}$-charged chiral multiplets 
receive radiative corrections at the one-loop level, 
and have masses at least 
of the order of $\sqrt{\alpha_{QCD}} m$.
They are pseudo-Nambu--Goldstone bosons and cannot remain 
exactly massless, because the SU(3)$_C$ interaction explicitly 
breaks the corresponding chiral global symmetry. 
One of the singlet pseudo-scalars also has a mass of the order of $m$ 
due to the mixed anomaly with SU(5)$_H$
(and an explicit breaking of an $R$ symmetry).
Only one pseudo-scalar field remains massless below the 
supersymmetry-breaking scale, and that field plays 
the role of the QCD axion.

Although there exists an accidental Peccei--Quinn symmetry
as advocated above,
higher-dimensional operators such as
\begin{eqnarray}
 \int_{y=0} d^4x d^2\theta d^2\bar{\theta} {z_0 \over M^4}(Q)^3QD^2Q
 +\int_{y=l} d^4x d^2\theta d^2\bar{\theta} {z_l \over M^4}
 ({\bar Q})^3{\bar Q}D^2{\bar Q}
 \label{eq:breaking}
\end{eqnarray}
break it explicitly,
where $z_0$ and $z_l$ are dimensionless coupling constants 
of order one,
$D$ denotes the covariant
superderivative, and the implicit gauge indices are
contracted so that the terms are gauge invariant.\footnote{The 
operators in (\ref{eq:breaking}) are not necessarily 
gauge invariant and are absent, when the extra quarks have 
non-trivial U(1)$_Y$ charges. 
The extra potential of the axion due to the explicit-breaking 
operators can be suppressed more if it is the case.}

Let us make a conservative estimate
of the QCD axion effective potential
\cite{Din}
induced by the explicit breaking operators 
localized at the fixed points.
On dimensional grounds,
the dominant contributions
to the axion potential
turn out to be of order
\begin{equation}
 {m^3 \over F_{PQ}^{20}} {\Lambda_H^{11} F_{PQ}^8}
 {z_0^* \over M^4} {z_l^* \over M^4} F_{PQ}^{10}
 \sim m^4\left({F_{PQ} \over M}\right)^8,
\end{equation}
where the spurion charges of $\Lambda_H^{11}$ for the selection rules
are apparent from Eq.(\ref{eq:nonpert})
and the supersymmetry-breaking scale factor $m^3$ originates from the
superderivatives and superspace integrals.
These corrections should not be too large to make 
the Peccei--Quinn mechanism ineffective:
\begin{equation}
\left[ m^4 \left(\frac{F_{PQ}}{M}\right)^8
 = 10^{-24} \GEV^4 \times 
 \left(\frac{m}{10^4 \GEV}\right)^4 
 \left(\frac{F_{PQ}}{M}10^5\right)^8 \right]
 \lsim 10^{-10} \times \frac{m_u m_d}{m_u+m_d}\Lambda^3_{\chi SB},
\label{eq:suffice}
\end{equation}
where $m_u$ and $m_d$ denote the masses of the up and down quarks, and 
$\Lambda_{\chi SB}$ is the energy scale of the QCD chiral 
symmetry breaking.

\section{Cosmological Issue}

Owing to the dynamical condensation
$\langle Q{\bar Q} \rangle \sim F_{PQ}^2$,
there exist colored particles in the pseudo-Nambu--Goldstone 
multiplets in the low-energy spectrum, as mentioned above.
These particles, if they were to live too long, 
would constitute dark matter
and lead to a cosmological difficulty \cite{Dark}.

We allow\footnote{The hypercharge should be assigned 
to the extra-quarks appropriately.
We do not go into the arguments on the gravitational
anomaly cancellation in this paper.}
 such gauge-invariant terms as
\begin{eqnarray}
 \int_{y=0} d^4x d^2\theta d^2\bar{\theta} {z \over M}Q^*Q{\bar d},
 \label{eq:decayop}
\end{eqnarray}
where $z$ is a dimensionless coupling constant of order one,
${\bar d}$ denotes the down-quark chiral superfield at $y=0$
(see footnote 4),
and the implicit gauge indices are contracted.
In the presence of this operator, 
all the colored particles have sufficiently short lifetime.

We note that the presence of the above terms does not alter
the conclusion in the previous section.
In fact, the operator (\ref{eq:decayop}) does not 
contribute to the extra potential of the axion, as seen  
in the same analysis as that at the end of section 3 .

\section{Flow of Gauge Coupling Constants}

\begin{figure}[h]
 \begin{center}
\begin{picture}(250,200)(50,0)
   \resizebox{12cm}{!}{\includegraphics{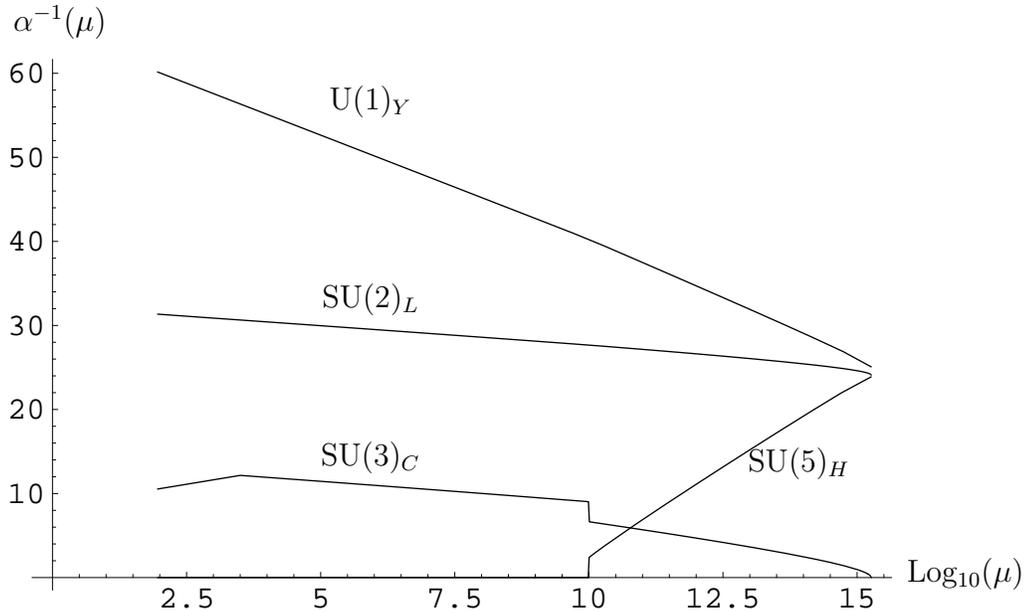}} 

   \Text(-200,60)[c]{SU(3)$_C$}
   \Text(-200,120)[c]{SU(2)$_L$}
   \Text(-200,195)[c]{U(1)$_Y$}
   \Text(-38,58)[c]{SU(5)$_H$}

   \Text(25,10)[b]{Log$_{10}(\mu)$}
   \Text(-300,225)[r]{$\alpha^{-1}(\mu)$}
\end{picture}
 \caption{Renormalization-group flow of the gauge coupling constants of
  the model. The gauge-coupling constants of the MSSM at 91GeV 
  contain threshold corrections from a typical spectrum of 
  the MSSM particles. The averaged mass of the extra particles, 
  some of which are charged under SU(3)$_{C}$ and U(1)$_Y$, 
  are set to be $10^{3.5}$GeV. 
 The Peccei--Quinn scale $F_{PQ}$ is set to $10^{10}$GeV. 
 The renormalization group is based on a U(1)$_Y$ charge assignment
  under which $Q^i{}_\alpha$ and $\bar{Q}_i{}^\alpha$ are 
  neutral, for definiteness. 
Two-loop effects have been taken into account.}
\label{fig:rgflow}
 \end{center}
\end{figure}

The colored extra multiplets in the low-energy spectrum has 
impacts not only in cosmology but also in the renormalization-group 
flow of gauge coupling constants.
In particular, the SU(3)$_C$ coupling becomes asymptotically 
non-free, 
and the model described in section \ref{sec:model} 
no longer serves as a good description at very short distance scale, 
since the QCD interaction becomes non-perturbative.
A typical renormalization-group flow of the gauge-coupling constants 
is shown in Fig.~\ref{fig:rgflow}.%
\footnote{The running of the QCD coupling in the high-energy regime
is insensitive to the Peccei--Quinn scale at the one-loop level.} 
It shows that the Kaluza--Klein scale 
can be as high as $10^{15}$ GeV, and the cutoff scale 
as high as $10^{16}$ GeV, for $m \simeq 10^{3.5}$ GeV.
Thus, some of the extra QCD-charged particles 
can be within the reach of LHC, while 
sufficient suppression 
$F_{PQ}/M\lsim 10^{-5}$ is obtained in Eq.(\ref{eq:suffice}),
even if the terms Eq.(\ref{eq:breaking}) are not forbidden by 
the U(1)$_Y$ charge.

\section*{Acknowledgements}

I.~K.-I.~would like to thank T.~Kugo and K.~Ohashi
for valuable discussions.
This work is partially supported by 
the Japan Society for the Promotion of Science, 
the Miller Institute for the Basic Research of Science, 
the Director, Office of Science, Office of High Energy and 
Nuclear Physics, of the U.S. Department of Energy 
under Contract DE-AC03-76SF00098 (T.W.), 
and Grant-in-Aid Scientific Research (s) 14102004 (T.Y.).

\appendix

\section{Kink Configuration and Supersymmetry}

It is shown in this appendix that the kink configuration of the 
coefficient of the Chern--Simons interaction in Eqs.(\ref{eq:CS-U(1)}) 
and (\ref{eq:kink}) can be understood as a VEV of a background field. 
The origin of the kink solution is also explained.

In six-dimensional spacetime, the interaction 
\begin{equation}
  S_{CS} = - \int G^{(1)} \wedge I^{(5)},  
\label{eq:WZ-a}
\end{equation}
is consistent with the minimal SUSY, where $G^{(1)}$ is 
1-form field strength of a scalar field $C^{(0)}$
in a linear hypermultiplet, and $I^{(5)}$ a 5-form that satisfies 
\begin{equation}
\tr F^3 = d I^{(5)}; 
\end{equation}
$F$ is the field strength of a Yang--Mills field.
This interaction is known in the context of string theories
as a part of the Wess--Zumino interaction
\begin{equation}
 \int C^{(p+1)}- G \wedge I_{(0)}
\end{equation}
on D$p$-brane world volumes,
where $C^{(p+1)}$ is the Ramond--Ramond $(p+1)$-form potential, 
$G$ a collection of Ramond--Ramond field strengths, and 
$I_{(0)}$ a differential form that is related with the anomaly
polynomial $I$ through $I = d I_{(0)}$ \cite{Dou}.

When there is a magnetic source of the field $C^{(0)}$, 
its Bianchi identity is given by
\begin{equation}
d G^{(1)} = \sum_i N_i \delta^{(2)}({\bf y}-{\bf y_i}),
\end{equation}
where
$\delta^{(2)}$ denotes a 2-form supported only on a point ${\bf y_i}$
(of a magnetic source), and $N_i$ the magnetic charge located there.

The interaction (\ref{eq:WZ-a}) implements the inflow of anomaly:
Let us introduce a 4-form $I^{(4)}$
which satisfy 
\begin{equation}
 \delta_\epsilon I^{(5)} = d I^{(4)},
\end{equation}
where $\delta_\epsilon$ denotes the gauge variation.
Then the variation of the action (\ref{eq:WZ-a}) is given by
\begin{equation}
\delta_\epsilon S_{CS}
= - \int G^{(1)} \wedge d I^{(4)} 
= \sum_i N_i \int_{{\bf y}={\bf y_i}} I^{(4)},
\end{equation}
and hence the triangle anomaly flows into a singularity
by the amount proportional to the charge localized there.

Let us assume that the internal space of the two extra dimensions 
is $T^2/Z_2$.
One can consider a torus which is long in one direction, and 
short in the other.
Then, one has an effective description in five
dimensions, 
by performing Kaluza--Klein reduction in the short direction.
This five-dimensional description is what we need in the main text.

Let us suppose that ${\bf y} = (y,z)=(0,0)$ singularity 
has a unit magnetic charge 
of $C^{(0)}$, and the $(y,z)=(l,0)$ singularity has the opposite charge. 
Then, the 1-form $G^{(1)}=dC^{(0)}$ has a positive  rotation
and a negative one, 
respectively, around those singularities.
The Hodge dual of the 1-form $G^{(1)}$ is given by a 5-form 
\begin{equation}
*(G^{(1)}) = {\tilde G} \wedge \epsilon_{\mu\nu\kappa\lambda}
dx^\mu dx^\nu dx^\kappa dx^\lambda,
\end{equation}
with ${\tilde G}$ a 1-form on $T^2/Z_2$ 
that has a positive  divergence and a negative one
at the singularities.
Then, the interaction (\ref{eq:WZ-a}) is rewritten as 
\begin{equation}
S_{CS} = - \int G^{(1)} \wedge I^{(5)} 
       = - \int d^6x \vev{*(G^{(1)}) ,I^{(5)}}
  \supset - \int d^5x \left(I^{(5)}\right)^{\mu\nu\kappa\lambda 4}   
\epsilon_{\mu\nu\kappa\lambda}{\tilde G}_4.
\end{equation}
In the limit where the second extra dimension is small, 
${\tilde G}_4$ is constant along $y \in (-l,0)$
as well as along $y \in (0,l)$.
Thus, the kink configuration $h(y)$ is given by ${\tilde G}_4$.

\end{document}